\documentclass[%
 reprint,
 showpacs,
 amsmath,amssymb,
 aps,
]{revtex4-1}

\usepackage{epsfig,graphicx}
\usepackage{dcolumn}
\usepackage{bm}

\begin{document}

\title{\bf Experimental test of the time stability of the half-life of alpha-decay $^{214}$Po nuclei}

\author{E.N.~Alexeyev}
\email{alexeyev@ms1.inr.ac.ru}
\author{Ju.M.~Gavriljuk}
\author{A.M.~Gangapshev}
\author{A.M.~Gezhaev}
\author{V.V.~Kazalov}
\author{V.V.~Kuzminov}
\email{bno\_vvk@mail.ru}
\affiliation{Baksan Neutrino Observatory INR RAS, Russia}
\author{S.I.~Panasenko}
\author{S.S.~Ratkevich}
\email{ratkevich@univer.kharkov.ua}
\affiliation{V.N.Karazin Kharkiv National University, Ukraine}
\author{S.P. Yakimenko}
\affiliation{Baksan Neutrino Observatory INR RAS, Russia}

\begin{abstract}
A method and results of an experimental test of the
time stability of the half-life of alpha-decay
$^{214}$Po nuclei are presented. Two underground
installations aimed at monitoring the time stability
have been constructed. Time of measurement exceeds
1038 days for one set up and 562 days for the other.
It was found that amplitude of  possible annual
variation of $^{214}$Po half-life does not exceed
0.2\% of the mean value. The limit on the deviation
of the decay curve from exponent at $0.034\cdot
T_{1/2}<t< 0.1\cdot T_{1/2}$ range has been found.
\end{abstract}


\date{May 26, 2011}
\maketitle


\section{Introduction}

A number of investigations aimed to check absolute
stability of half-life of different radioactive
isotopes and to search for possible time variations
of the half-life constants under action of known and
possible unknown natural factors has been carried
out during last years.

In work \cite{r1} it was shown that, after many
years of investigations with scintillation and
semi-conductor $\alpha$-ray detectors, a conclusion
was made about  changes of rate of radioactive
elements' $\beta$-decay with 24 hours and 27 days
periodicities.

Many years measurement's decay rates data for the $^{32}$Si ($\beta$-decay)
measured in the Brookhaven National Laboratory (USA, 1982-1986 years) and
for the
$^{226}$Ra ($\alpha-$ and $\beta$-decays) measured in the
Physikalisch-Technische-Bundesanstalt (Germany, 1983-1998 years) were
analyzed in work \cite{r2}, \cite{r3}, \cite{r4}.
Decay rate variations with a one year period and maximum amplitude of
$\sim$0.15\% in January-February were found in both data sets. The authors
considered an assumed seasonal variations of the detector systems
characteristics and/or the direct annular modulation of the count rate being
caused by some unknown factor depending on the Sun-Earth distance as
possible causes of such count rate variations.

The weak point of the count rate long time monitor experiments is
a possible influence of meteorological, climatic and geophysical factors on
the count rate of a source-detector pair.
This shortcoming could be practically totally avoided in measurements based
on a direct registration of a nuclear life time between birth and decay.
Such a method allows us to answer the question of possible change with time
of a just nuclear decay constant.

Besides that a direct registration of nuclear life time allows us to
study of the exponentially decay low. Some theoretical models
\cite{r5}, \cite{r6} predicted that decay curve does not exactly
follow an exponential low in the short-and long-time regions in
particular due to of the so-called quantum Zeno effect
 \cite{r7}, \cite{r8}, \cite{r9}, \cite{r10}. Experimentally Zeno effect was
proved \cite{r11} in repeatedly measured two-level system undergoing Rabi
transitions, but not observed in spontaneous decays.

Very important condition for measure possible deviations from an exponential
low is the all investigated nuclei could have the same age.

\section{Experimental method}

A primary task of the investigation presented in this work was to
investigate constancy of a half-life ${\bf {\tau}}$ $({\bf {\tau}}\equiv T_{1/2})$
of $^{214}$Po during several years. $^{214}$Po decays with 164.3 $\mu$s
half-life \cite{r12} by emitting the 7.687 MeV $\alpha$-particle. This
isotope appears mainly in the exited state ($\sim87$\%) in the $^{214}$Bi
$\beta$-decay. Half-lives of the exited levels does not exceed 0.2 ps
 \cite{r13}
and they discharge instantly with regard to the scale of the $^{214}$Po
half-life. Energies of the most intensive $\gamma$-lines are equal to 609.3 keV
(46.1\% per decay), 1120 keV (15.0\%) and 1765 keV (15.9\%). So, the $\beta$-particle
and $\gamma$-quantum emitted at the moment of a birth of $^{214}$Po nuclear (start)
and the $\alpha$-particle emitted at the decay moment (stop). Measurement
of "start-stop" time intervals allows one to construct decay curve at an
observation time and to determine the half-life from it's shape.

$^{214}$Bi isotope is an intermediate daughter product in the $^{238}$U
series. It gives almost all $\gamma$-activity of the series. This isotope
could be produced with a constant velocity if an intermediate isotope
$^{226}$Ra ($T_{1/2} = 1600$ y) of the $^{238}$U series will be used as
a source. A time of an equilibrium reached in a partial $^{226}$Ra-$^{214}$Bi
series equals to $\sim20$ days and is determined by the longest half-life
isotope, $^{222}$Rn (3.82 days). A $^{214}$Bi decay rate doesn't change
after this time if a source is hermetically sealed to prevent an escape of the radon.

Two test sources with $^{226}$Ra activity of $\sim90$ Bq and $\sim20$ Bq
 were prepared in the V.G.~Khlopin Radium Institute (St.Petersburg, Russia) in March, 2008.

A thin transparent radium spot is deposited in the center of a polished
plastic scintillator disc of 18 mm diameter and 0.8 mm thickness.
A disc's side with the spot is covered by a similar disc hermetically
glued on periphery (2.5 mm width). A plastic scintillator (PS) registers
all charged particles generated in the decay series. A PS light yield for
an $\alpha$-particle absorption is $\sim0.1$ of that for the electron
of the same energy. Due to this fact $\alpha$-spectra mix with
$\beta$-spectra if a scintillator is thick enough to absorb a complete
energy of an electron. To prevent this effect and separate these spectra
the scintillator discs were made thin enough so that electrons lose only
part of their energy.

\begin{figure}
\begin{center}
\includegraphics*[width=8.3cm]{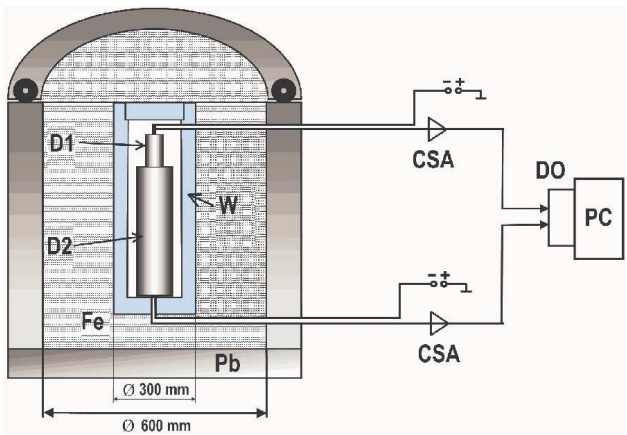}
\caption{\label{fig1} Schematic cross-sectional view of TAU-1 setup and electronics block diagram.}
\end{center}
\end{figure}

Two setups named TAU-1 and TAU-2 were made to measure radiations
of these sources. A schematic sectional view of the TAU-1 with the
electronics is shown in Fig.\ref{fig1}.
The detector D1 contains the photomultiplier (FEU-85) which views the source
disc through a plastic light guide. A teflon reflector is placed under the source
to improve light collection. The whole assembly is packed in a stainless steel body.
The detector D1 is placed on the butt-end of the scintillation detector D2 with
low background NaI(Tl) crystal of 80 mm diameter and 160 mm length. The crystal has a stainless still cover and a quartz entrance window. The crystal with photomultiplier (FEU-110) is packed in a cooper body. The two detectors are mounted vertically in a low background shield made from Pb(10 cm) + Fe(15 cm) + W(3 cm). The TAU-1 is located in the underground laboratory "KAPRIZ" of the Baksan Neutrino Observatory of INR RAS at 1000 m of water equivalent depth. A cosmic ray background is decreased by $\sim10^4$ times due to the rock thickness in comparison with the ground surface one. Walls of the laboratory are covered by a low background concrete thus decreasing by $\sim8$ times the $\gamma$-ray background from a rock natural radioactivity. Total background suppression inside the shield is equal to $\sim2000$ times.

Signals from the photomultipliers are read by charge sensitive preamplifiers
(CSA) and fed to two inputs of a digitizer (digital oscilloscope - DO)
LA-n20-12PCI which is inserted into a personal computer (PC). Pulses are
digitized with 6.25 MHz frequency. The DO pulse recording starts by a signal
from the D2 which registered $^{214}$Bi decay's $\gamma$-quanta. A D2 signal
opens a record of a sequence with 655.36 $\mu$s total duration where first
81.92 $\mu$s time is a "prehistory" and the last 573.44 $\mu$s is a "history".
Duration of a "history" exceeds the three $^{214}$Po half-lives.

A PS disc source in the TAU-2 installation is fixed at the end of an air
light guide having a smooth wall made of VM-2000 light reflecting film.
The light guide is put into 0.8 mm thick stainless still rectangular
frame with inner dimensions of $150\cdot23\cdot9$ mm. An open butt-end of
the frame is welded into a bottom of cylindrical stainless still body of
45 mm diameter and 165 mm length where photomultiplier (FEU-85) placed.
So much for the detector D1 of the TAU-2. Two scintillation detectors
(D2a and D2b) made of large NaI(Tl) crystals ($150\cdot150$ mm) are used
for a registration of the $\gamma$-quanta. Each crystal has a stainless
still cover and a quartz entrance window.
Photomultipliers (FEU-49) are used for the light registration. The D1 are
installed into a gap between D2a and D2b. A scheme of pulse registration in
the TAU-2 is similar to the one of TAU-1 but signals from D1a and D2b are
preliminarily summed in additional summer. The TAU-2 is located in the low
background room in the deepest underground laboratory NLGZ-4900 of the BNO
INR RAS at the 4900 m of water equivalent depth. A cosmic ray background is
decreased by $\sim~10^7$ times by the rock thickness in comparison with the
ground surface one. The room walls are made of polyethylene (25 cm) +
Cd(0.1 cm) + Pb(15 cm). The detectors are surrounded with additional lead
shield of 15 cm thickness aimed to absorb a radiation from decays of
$^{214}$Bi in the air being in equilibrium with the radon in the room.

\section{Working characteristics}

Working characteristics of each installation were tested preliminary by using usual multichannel analyzer AI-1024.

\emph{\textbf{TAU-1 installation}}

\begin{figure}
\includegraphics*[width=8.5cm]{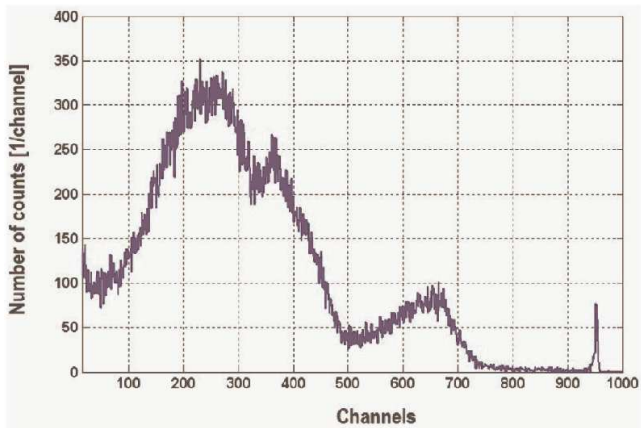}
\caption{\label{fig2} D1 detector pulse amplitude spectrum collected at 600 s.}
\end{figure}

An amplitude spectrum, from the D1, collected during 600 s is shown in Fig.\ref{fig2}.
\begin{figure}
\includegraphics*[width=8.5cm]{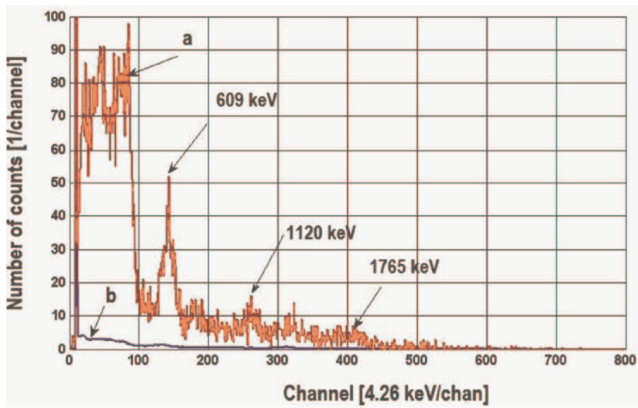}
\caption{\label{fig3} D2 detector pulse amplitude spectra (a) of $^{226}$Ra source (30 min collection time ) and (b) of background normalized at 30 min (72 hours collection time).}
\end{figure}
Integral count rate is equal to 179 s$^{-1}$. A wide peak above the 500 channel corresponds to $^{214}$Po $\alpha$-particles. A count rate within $500\div750$ channels is equal to 21.2 s$^{-1}$. An amplitude spectrum, from the D2, collected during 30 min is shown in Fig.\ref{fig3}a. A main part of the spectrum is due to $\gamma$-quanta from the source. A spectrum of the D2 background was measured in 72 hours when the D1 was removed from the shield. It's spectrum shown in the Fig.3b. An integral count rate in the spectrum in Fig.\ref{fig3}a is 4.90 s$^{-1}$ (1.64 s$^{-1}$ at the energy above 400 keV) and that one in Fig.\ref{fig3}b is 0.20 s$^{-1}$.

\emph{\textbf{TAU-2 installation}}

Integral count rate of the D1 is equal to 70 s$^{-1}$. A background count rate of the D2a detector at the energy above 400 keV is equal to 2.3 s$^{-1}$. A total count rate with the source is equal to 5.0 s$^{-1}$. A light collection coefficient of the D1 in the TAU-2 was found to be 0.51 of the one in the TAU-1.

\section{Useful event selection}

The TAU-1 and TAU-2 installations were placed into underground low background shields to provide stable environmental conditions and a low cosmic ray background. It allows us to decrease possible variations of electronics characteristics and to exclude random D1-D2 coincidences caused by D2 registrations of background's muons and $\gamma$-quanta.

\begin{figure}
\includegraphics*[width=7.5cm]{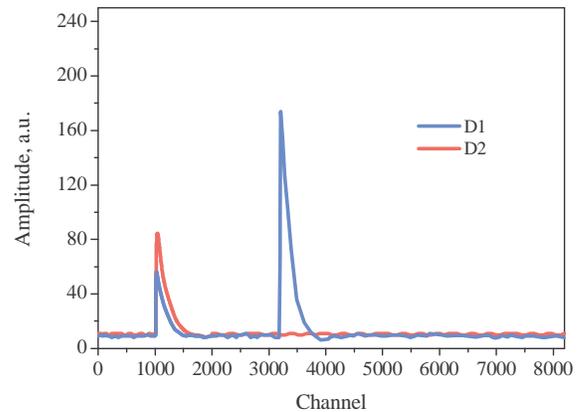}
\caption{\label{fig4} An example of coinciding D1-D2 events with two pulses in the D1 "history" (blue) and one pulse in the D2 "history" (red).}
\end{figure}

\begin{figure}
\includegraphics*[width=8cm,angle=0.]{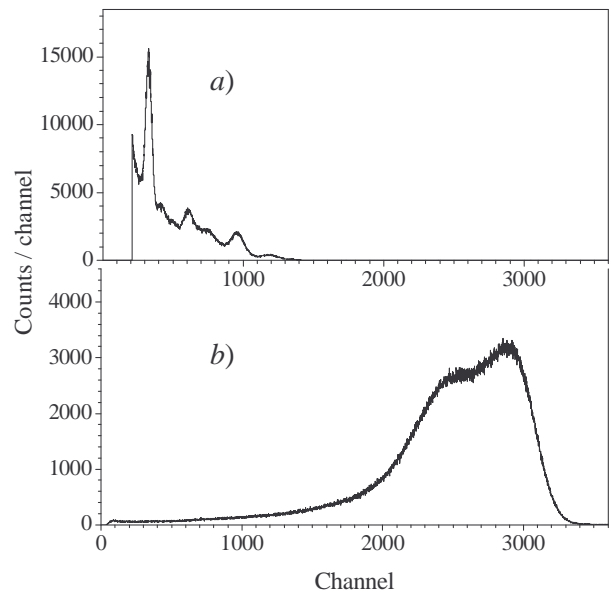}
\caption{\label{fig5} Spectra of the D2 $\gamma$-quantum pulses (a) and the second D1 $\alpha$-particle pulses (b)
of the TAU-1 setup.}
\end{figure}

A control of the DO work in regimes of on-line event separation,
data collection and visualization
was carried out by specially designed PC program. This program
allows DO to record only coinciding D1-D2 events, with two pulses
in the D1 "history" and one pulse in the D2 "history". The last one
is in prompt coincidence with the first D1 pulse. An example of
the event is shown in Fig.\ref{fig4}.
A number of such events make
up $\sim48$\% of the total amount. Their count rates are equal to $\sim4.3$ s$^{-1}$ for TAU-1 and $\sim4.6$ s$^{-1}$
for TAU-2 for the energy threshold of 400 keV in the channels of D2 detector.
Spectra of the D2 pulses ($\gamma$-quanta) and the second D1 pulses ($\alpha$-particles) for the TAU-1 are shown in Fig.\ref{fig5}(a,b).
Double peak of the Fig.\ref{fig5}b spectrum can be explained by different light output of the two PS disks. The lower energy peak was found to move towards the beginning
of the coordinates with time. This effect could be connected with a degradation of PS surface scintillations properties under the active spot due to diffusion of chemical components or decay products into a surface layer of the source deposited disc.

\section{On-line data processing}

A rate of the direct data record is 5.8 Gb$\cdot$day$^{-1}$. Such large information flux complicates
processes of the data collection and off-line processing. A number of pulses and their delays for each event are analysed on-line to decrease usage of PC memory. "Wrong" events are excluded.
Only a time of appearing and amplitude of pulses defined for the "right" events record at the PC memory. A rate of an information accumulation decreased up to 8.5 Mb$\cdot$day$^{-1}$.
Measurements with the TAU-1 setup started in April, 2008 and the ones with TAU-2 started in July, 2009.

\section{Results and discussion}

A continuous data set is divided on a sequence of equal intervals. A spectrum of delays between the first and second D1 pulses is constructed for each interval. This spectrum is approximated by exponential function viewed as $y=a\cdot exp(-ln(2)\cdot t/\tau)+b$ corresponding to the decay curve. A value of the half-life $\tau$ is found from this curve. The series of $\tau$-values is analysed to find out any regularity or variation.

\begin{figure}
\includegraphics*[width=7.5cm,angle=0.]{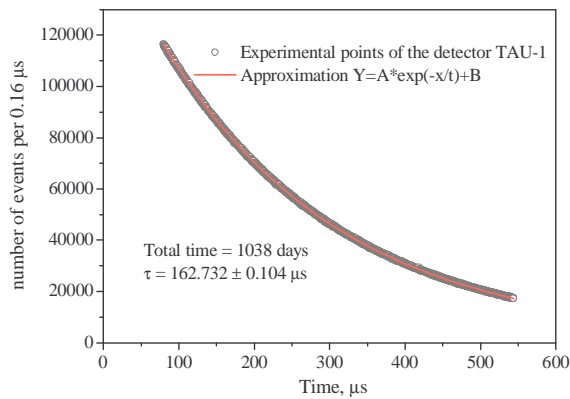}
\caption{\label{fig6} Summary exponent constructed for the whole data set collected at 1038 days by TAU-1.}
\end{figure}

Duration of the analyzed data sets of TAU-1 and TAU-2 were equal to 1038 days and 562 days correspondingly. A half-life value obtained for a decay curve constructed with the whole set of TAU-1 data was found to be $(162.733 \pm 0.104)$ $\mu$s. This curve is shown in Fig.\ref{fig6}.
A half-life value
obtained for the TAU-2 data was found to be $(164.249 \pm 0.115)$ $\mu$s.
The difference between values is
connected with uncertainties of the digitizer's frequency calibrations.

\begin{figure}
\includegraphics*[width=7.5cm,angle=0.]{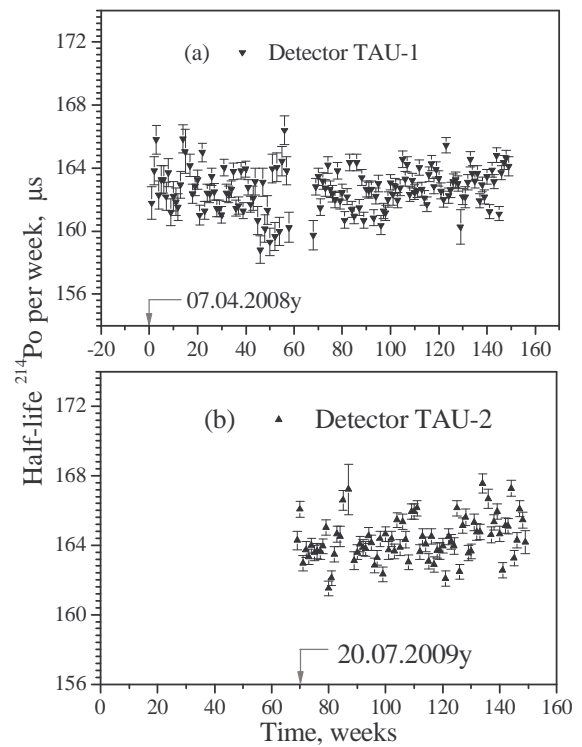}
\caption{\label{fig7} Time dependence of $^{214}$Po half-life determined for
 one week collection interval
data set from TAU-1 (a) and TAU-2 (b). }
\end{figure}

Each interval of the whole
data set was chosen to be 7 days for the analysis of the half-life constant
behavior for a long time period.
Sequences of $\tau$-values for TAU-1 and TAU-2 are shown in
Fig.\ref{fig7} (a)
and (b) respectively.
\begin{figure}
\includegraphics*[width=7.5cm,angle=0.]{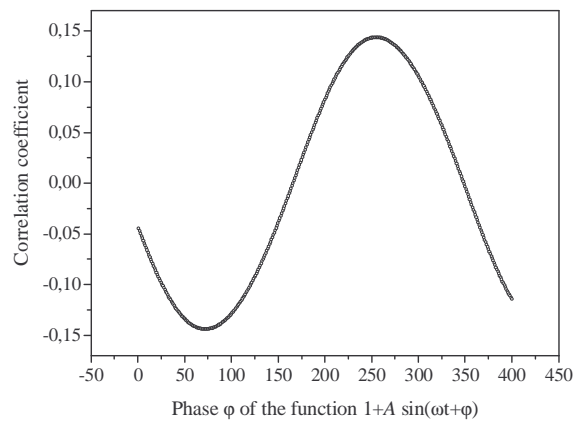}
\caption{\label{fig8} Dependence of $k(\varphi)$ for
the TAU-1 data.}
\end{figure}
To search for a possible annular variations of the data they were
normalized on the averaged values and compared with a periodical
function $f(t) = [1+ A \cdot sin(\omega t+\varphi)]$ where $\omega =
1/365$ day$^{-1}$, $A$ and $\varphi$ are the amplitude and phase
correspondingly. $A = 0.002$ was taken at the beginning. A
$\varphi$-value has been changed from 1 to 365 with the step of 1
day. A correlation coefficient $k$ between the $\tau$-sequences and
$f(t)$ was calculated for the each $\varphi$-value. A dependence of
$k(\varphi)$ for the TAU-1 data is shown in Fig.\ref{fig8}.
\begin{figure}[pt]
\includegraphics*[width=7.0cm,angle=0.]{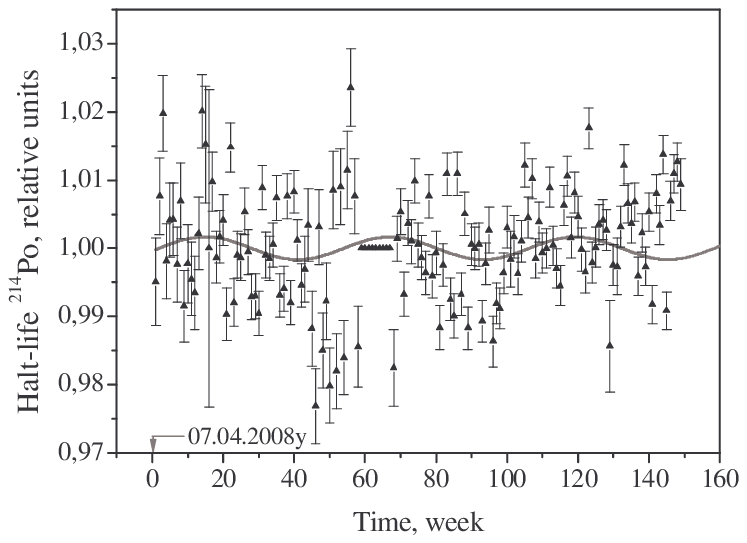}
\caption{\label{fig9} Normalized graph of the time
dependence of $^{214}$Po half-life for TAU-1 in
comparison with function $f(t) = [1+0.00166 \cdot
sin(t/365+255)]$, $t$ - days.}
\includegraphics*[width=7.0cm,angle=0.]{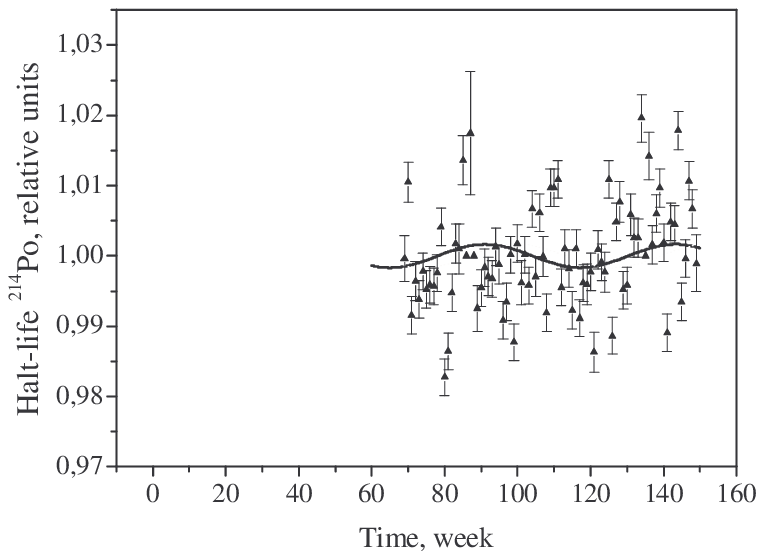}
\caption{\label{fig10} Normalized graph of the time
dependence of $^{214}$Po half-life for TAU-2 in
comparison with function $f(t) = [1+0.00166 \cdot
sin(t/365+88)]$, $t$ - days.}
\end{figure}

The maximum value $k=0.144$ has been reached at $\varphi = 225$. A selection
of $A$-value corresponding to the $\chi^2$ minimum was done for this
$\varphi$-value, and was found to be $A = 0.00166$. The values of
$k = 0.160$, $A = 0.00166$ and $\varphi = 88$ days were found for TAU-2 by
the similar way.

Corresponding graphs for the TAU-1 and TAU-2 data are shown in
Fig.\ref{fig9} and Fig.\ref{fig10}.
The estimated values of $A$ are the same for both setups but the phases
differ by 167 days.
A mean square error $s(k)$ of a $k$ for a number $N$ of degrees of freedom
could be estimated as
$s(k) \approx (1-k^2)/\sqrt{N}$. It was found that $s(k)=0.080$ for the
TAU-1 ($k=0.144$ and $N=149$) and $s(k)=0.109$ for the TAU-2
($k=0.160$ and $N=81$).

\begin{figure}[pt]
\includegraphics*[width=7.0cm,angle=0.]{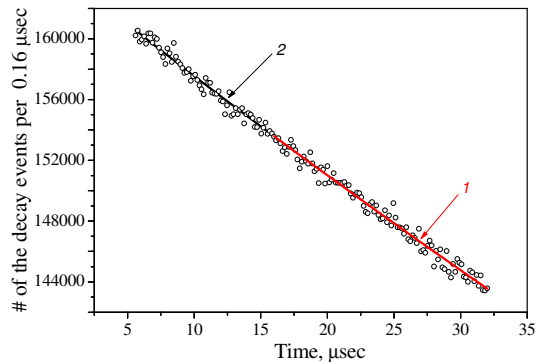}
\caption{\label{fig12} The decay dependence of
$^{214}$Po of the detector TAU-2. Points is the
experimental data. Curve \emph{1} is the result of
the approximation of the exponential low. 
Curve \emph{2} is the result of
the aproximation using eq.1}
\end{figure}

A low reliability of the obtained results and disagreement between
$\varphi$-values allow one to determine only the upper limit
of the amplitude of $\tau$-constant possible annular variation which
was found to be $A \leq 0.002$ (90\% C.L.). An interval of data spacing
was chosen to be 1 day in order to search for possible short period
variations. No significant frequency peaks in a range
$0.003 \div 1.000$ day$^{-1}$ of the Fourier spectra constructed by using
two such $T_{1/2}$ - series has been found.

We have tested the theoretical consideration of possible
violation of the radioactive decay exponential law at low values of time
of life $(t \lesssim 0.1 \tau)$ of $^{214}$Po nucleus. For this purpose
the total data set of the detector TAU-2 was used. The exponent was
constructed for the time interval $5.6 \div 544$ $\mu$sec. Half-live of the
$^{214}$Po nucleus was determined using data for the time interval $16\div544$
$\mu$sec. Then the residual part of the data was used for analysis of the
deviations between the data and the exponent which was calculated.

The deviation was search for in form \cite{r5}:
\begin{equation}
\label{eq1}
exp(-t/\tau)+\sum_{i=1}^na_{i}(t/\tau)^{-n}.
\end{equation}
Limiting n=2 we find:
a$_{1}$ $\leqslant$ 4.01$\cdot$10$^{-6}$,  a$_{2}$ $\leqslant$ 3$\cdot$10$^{-7}$,
for time interval $(0.034 \cdot T_{1/2} \leqslant t \leqslant  0.1 \cdot T_{1/2})$.

The results of the analysis are shown at the fig.11. 
The points at the figure are the
experimental data for the time interval $5.6 \div 32$ 
$\mu$sec. Curve \emph{1} is the part of the exponent
calculated for time interval $16 \div 544$ $\mu$sec.
Curve \emph{2} is the result of the approximation of
the data for time interval $5.6 \div 16$ $\mu$sec, using
eq.(\ref{eq1}).

\section{Conclusions}

1. Two underground installations aimed at monitoring the time stability of  $^{214}$Po half-life have been constructed.

2. Time of measurement exceeds , at present, 1038 days for TAU-1 and 562 days  for TAU-2.

3. It has been found that amplitude of possible  annual variation of $^{214}$Po half-life  does not exceed 0.2\% of the mean value.

4. No deviation of the decay curve from exponent  at $0.034 \cdot T_{1/2} < t < 0.1 \cdot T_{1/2}$ has been  found with an accuracy of  $\sim \cdot10^{-6}$.


\begin{thebibliography}{11}

\bibitem{r1} Yu.A.~Baurov, Yu.G.~Sobolev, Yu.V.~Ryabov, V.F.~Kushniruk,
"Experimental Investigations of Changes in the Rate of Beta Decay of Radioactive Elements".
Phys.of Atomic Nucl. V.\textbf{70}, No.11, 1825  (2007).

\bibitem{r2}
Jere H. Jenkins at al.,
"Evidence for Correlations Between Nuclear Decay Rates and Earth-Sun
Distance". Astropart. Phys. \textbf{32}, 42  (2010).

\bibitem{r3}
Peter A. Sturrock at al., "Concerning the Phases of Annual Variations of
Nuclear Decay Rates". Astrophysical Journal \textbf{737}, 65 (2011).

\bibitem{r4}
Jere H. Jenkins at al., "Analysis of Experiments Exhibiting Time-Varying
Nuclear Decay Rates: Systematic Effects or New Physics?". arXiv:1106.1678 [nucl-ex];
"Evidence for Time-Varying Nuclear Decay Rates: Experimental Results and
Their Implications for New Physics" arXiv:1106.1470 [nucl-ex].

\bibitem{r5} P.M.Gopych and I.I.Zaljubovsky, "Exponentiality of the basic law
of radioactive decay". PEPAN, \textbf{19}, 4, 785, (1988).

\bibitem{r6} L.~Fonda, G.C.~Ghirardi and A.~Rimini, "Decay theory of unstable quantum systems". Reports on Progress in
Phisics \textbf{41}, 587 (1978).

\bibitem{r7} L.A. Khalfin, "Zeno's quantum effect". Physics-Uspekhi \textbf{33}, 10, 868 (1990).

\bibitem{r8} B.~Misra and E.C.G.~Sudarshan, "The Zeno's paradox in quantum theory". J.Math.Phys. \textbf{18}, 756 (1977).

\bibitem{r9} P. Facchi and S. Pascasio, "Quantum Zeno dynamics: mathematical and physical aspects". J.Phys. A: Math.Teor. \textbf{41} (2008)
493001.

\bibitem{r10} Francesco Giacosa and Giuseppe Pagliara,"Oscillations in the decay law: a possible quantum mechanical explanation of the GSI anomaly". arXiv:1110.1669 [nucl-th].

\bibitem{r11} Wayne M. Itano, D. J. Heinzen, J. J. Bollinger, and D. J. Wineland, "Quantum Zeno effect".
Phys.Rev. A\textbf{41}, 2295 (1990).

\bibitem{r12} E.~Browne, Nuclear Data Sheets \textbf{99}, 649 (2003).

\bibitem{r13} Table of Isotopes, Edited by R.B. Firestone et al., 8th ed.,
(Willey, New York, 1996).

\bibitem{r14} Physical Quantaty Data-book. (in Russian). Under I.K.Kikoin edition.
M. Atomizdat. 1976.

\end{thebibliography}
\end{document}